\documentclass[journal=nalefd,manuscript=article]{achemso}

\usepackage[utf8]{inputenc}
\usepackage{amsmath, amssymb}
\usepackage{amsfonts}
\usepackage{amssymb}
\usepackage[version=3]{mhchem}
\usepackage[]{graphicx}
\usepackage[T1]{fontenc}
\usepackage{lmodern}
\usepackage{hyperref}
\usepackage{xcolor}
\usepackage{notes2bib}
\usepackage{verbatim}
\usepackage{geometry}
\usepackage{mciteplus}

\definecolor{linkcol}{rgb}{0,0,0.4}
\definecolor{citecol}{rgb}{0.5,0,0}

\hypersetup{
	bookmarksopen=true,
	pdftitle="Semiconducting van der Waals Interfaces as Artificial Semiconductors",
	pdfauthor="E.Ponomarev et al",
	pdfsubject="Semiconducting van der Waals Interfaces as Artificial Semiconductors", %subject of the document
	pdfstartview={FitH},    % fits the width of the page to the window
	pdfmenubar=true, %menubar shown
	pdfhighlight=/O, %effect of clicking on a link
	colorlinks=true, %couleurs sur les liens hypertextes
	pdfpagemode=UseNone, %aucun mode de page
	pdfpagelayout=SinglePage, %ouverture en simple page
	pdffitwindow=true, %pages ouvertes entierement dans toute la fenetre
	linkcolor=linkcol, %couleur des liens hypertextes internes
	citecolor=linkcol, %couleur des liens pour les citations
	urlcolor=blue
}

\author{Evgeniy Ponomarev}
	\affiliation{DQMP and GAP, Université de Genève, 24 quai Ernest Ansermet,\\ CH-1211, Geneva, Switzerland}
	
	\author{Nicolas Ubrig}
	\affiliation{DQMP and GAP, Université de Genève, 24 quai Ernest Ansermet,\\ CH-1211, Geneva, Switzerland}
	
	\author{Ignacio Gutiérrez-Lezama}
	\affiliation{DQMP and GAP, Université de Genève, 24 quai Ernest Ansermet,\\ CH-1211, Geneva, Switzerland}
	
	\author{\\Helmuth Berger}
	\affiliation{Institut de Physique de la Matière Complexe, EPFL, CH-1015 Lausanne, Switzerland}
		
	\author{Alberto F. Morpurgo}
	\email{Alberto.Morpurgo@unige.ch}
	\affiliation{DQMP and GAP, Université de Genève, 24 quai Ernest Ansermet,\\ CH-1211, Geneva, Switzerland}

	
\title{\texorpdfstring{\Large Semiconducting van der Waals Interfaces\\ as Artificial Semiconductors} {Semiconducting van der Waals Interfaces\\ as Artificial Semiconductors}}

\begin{document}

\begin{abstract}
Recent technical progress is demonstrating the possibility to stack together virtually any combination of atomically thin crystals of van der Waals bonded compounds to form new types of heterostructures and interfaces. As a result, there is the need to understand at a quantitative level how the interfacial properties are determined by the properties of the constituent 2D materials. We address this problem by studying the transport and opto-electronic response of two different interfaces based on transition metal dichalcogenide monolayers, namely WSe$_2$/MoSe$_2$ and WSe$_2$/MoS$_2$. By exploiting the spectroscopic capabilities of ionic liquid gated transistors, we show how the conduction and valence bands of the individual monolayers determine the bands of the interface, and we establish quantitatively -- directly from the measurements -- the energetic alignment of the bands in the different materials, as well as the magnitude of the interfacial band gap. Photoluminescence and photocurrent measurements allow us to conclude that the band gap of the WSe$_2$/MoSe$_2$ interface is direct in $k-$space, whereas the gap of WSe$_2$/MoS$_2$ is indirect. For WSe$_2$/MoSe$_2$ we detect the light emitted from the decay of interlayer excitons and determine experimentally their binding energy using the values of the interfacial band gap extracted from transport measurements. The technique that we employed to reach this conclusion demonstrates a rather general strategy to characterize quantitatively the interfacial properties in terms of the properties of the constituent atomic layers. The results presented here further illustrate how van der Waals interfaces of two distinct 2D semiconducting materials are composite systems that truly behave as \emph{artificial semiconductors}, whose properties can be deterministically defined by the selection of the appropriate constituent semiconducting monolayers.
\end{abstract}

The absence of covalent bonds between the layers of van der Waals (vdW) materials is essential not only to allow the exfoliation of monolayers of excellent electronic quality from bulk crystals, but also to give an unprecedented flexibility in employing these monolayers to assemble new types of heterostructures\cite{geim_van_2013,novoselov_2d_2016,iannaccone_quantum_2018}. It is because of the absence of covalent bonds that monolayers of different compounds can be stacked on top of each other without constraints imposed by the need to match their crystalline lattices or by chemistry compatibility. As a result, a very rich variety of building blocks -- including semiconductors\cite{splendiani_emerging_2010,mak_atomically_2010}, semimetals\cite{wang_tuning_2015}, topological insulators\cite{fei_edge_2017}, magnets\cite{Gong2017,huang_layer-dependent_2017}, superconductors\cite{Cao2015, xi_ising_2016}, and more -- can be readily combined together to create artificial systems that were impossible to realize until now. That is why 2D materials offer a truly unprecedented potential to discover new physical phenomena or to engineer novel electronic functionalities. First examples are provided by the emergence of minibands and of satellite Dirac points in graphene-on-hBN \cite{yankowitz_emergence_2012,dean_hofstadters_2013, ponomarenko_cloning_2013}, the possibility to induce strong spin-orbit interaction in systems of Dirac fermions in graphene-on-semiconducting transition metal dichalcogenides (TMDs)\cite{wang_strong_2015-1,Wang2016}, or the occurrence of superconductivity in magic-angle graphene bilayers \cite{cao_unconventional_2018}.\\

Despite the vast scope of possibilities enabled by vdW interfaces, a systematic microscopic understanding allowing the interfacial electronic properties to be predicted in terms of those of the constituent monolayers is missing. Developing such an understanding in general is difficult, because depending on the specific interface considered many different microscopic processes  --hybridization of the electronic states in the two monolayers, relaxation of the crystalline structure, interaction effects\cite{novoselov_2d_2016}, etc.-- can play a prime role. To progress at this stage it is useful to  focus our attention on an important class of systems for which we can exploit existing intuition, namely that of semiconducting monolayers forming an interface that also behaves as a semiconductor, whose properties can be deterministically controlled by selecting the two constituent semiconductors in the vast palette of existing 2D materials.\\

At the simplest level, the strategy is to choose the constituent monolayers with an appropriate band alignment, so that the conduction band of the interface is inherited from one of the monolayers and the valence band from the other (as it happens in so-called type II semiconducting heterostructures)\bibnote{The case of a band alignment resulting in either conduction or the valence bands of the two monolayers that are degenerate in energy is not considered here. Despite providing even more possibilities to control the interfacial properties, that case is more complex because the hybridization of the states in the degenerate bands of the two materials can not be neglected.}. Under these conditions, many key interfacial semiconducting properties are distinct from those of the constituents, but uniquely determined by them. These include the size of the band gap (defined by the energetic alignment of the bands in the two monolayers), whether the gap is direct or indirect in $k-$space (determined by appropriately selecting the crystalline lattices of the materials forming the interface), the joint density of states governing absorption and radiative processes (directly linked to the properties of the bands in both constituent monolayers), and more. The interface is therefore a composite system that possesses unique semiconducting properties defined at the assembly stage by the choice of the constituent monolayers, which fully determine the low-energy optoelectronic response of the system detected experimentally. In other words, the interface truly behaves as an artificial semiconductor whose properties --that determine the optoelectronic response-- are created by design.\\

Whereas over the last couple of years considerable experimental effort has focused on vdW interfaces of different semiconducting 2D materials\cite{hong_ultrafast_2014,rivera_observation_2015,wang_interlayer_2016,kim_observation_2017}, only a very limited amount of work has been done to probe transport and optical properties on a same structure, enabling a full understanding of the energetics of vdW interfaces. That is why developing the ability to probe the properties of vdW interfaces based on 2D semiconductors to determine quantitatively -- directly from experiments -- how these properties are related to those of the constituent 2D materials, is now a key priority. Here, we start addressing these issues by means of a systematic investigation of the transport and optoelectronic properties of two prototype vdW interfaces based on semiconducting TMD monolayers, namely WSe$_2$/MoSe$_2$\cite{rivera_observation_2015} and WSe$_2$/MoS$_2$\cite{Zhu2017}.\\

Transport investigations rely on ionic liquid gated field-effect transistors \cite{Panzer2005, Shimotani2006, Misra2007, xie_organic_2011} realized on structures that enable the individual monolayers, and their interface, to be measured separately on a same device. By exploiting the spectroscopic capabilities of ionic liquid gating\cite{Braga2012, Jo2014, Lezama2014, ponomarev_hole_2018}, these devices enable us to establish a direct relation between the bands of the vdW interface and those of the constituent monolayers, and to determine quantitatively and precisely the offsets between the bands of the different monolayers, as well as the magnitude of the interface band gap. To probe whether interband transitions in the vdW interface are direct or indirect in $k$-space we perform photoluminescence (PL) and photocurrent (PC) experiments. Radiative decay of interlayer excitons is observed in WSe$_2$/MoSe$_2$ but not in WSe$_2$/MoS$_2$, in agreement with the expectation that the interfacial interband transitions are direct in the first case and not in the second\cite{yu_anomalous_2015}. The binding energy of interlayer excitons in WSe$_2$/MoSe$_2$ can then be determined by directly comparing the frequency of the light detected in  PL with the value of the interface band gap extracted from transport experiments. All together, the results presented here show that vdW interfaces behave as artificial semiconductors whose response -- determined by the choice of the constituent materials -- is virtually indistinguishable from that of a naturally existing 2D semiconducting material.\\

The device configuration employed for our transport measurements -- essential to obtain the results discussed below -- is illustrated in Figure 1 for the MoS$_2$/WSe$_2$ system. The MoS$_2$ and WSe$_2$ monolayers are exfoliated onto a Si/SiO$_2$ substrate (see Figure 1a and 1b). Using a by-now conventional "pick-up" technique based on polymer stamps\cite{zomer_transfer_2011}, the WSe$_2$ monolayer is transferred onto the MoS$_2$ monolayer with the aid of a motorized manipulator stage under an optical microscope. Figures 1c and 1d show microscope images taken after the transfer and after having attached contacts. The final step is the application of the top ionic liquid, leading to a device structure such as the one represented schematically in Figure 1e (with the gate and reference electrodes also shown; see section S1 in Supporting Information for details of the fabrication process). The important aspect of this device geometry is that the different parts of the structure can be measured independently in a field-effect transistor configuration, while being directly connected to each other. This is key as it allows a direct quantitative comparison of relevant quantities measured in the different parts of the device.\\

\begin{figure}
	\centering
	\includegraphics[width=0.8\textwidth]{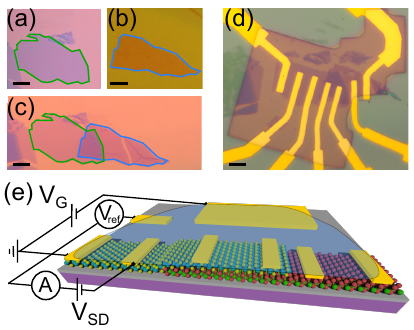}
	\caption{Device structure enabling the quantitative characterization of van der Waals interface and of their constituent monolayers. Optical microscope images of a MoS$_2$ (a), a WSe$_2$ (b) monolayer (delimited respectively by the green and the blue lines), and of the structure assembled by transferring the WSe$_2$ monolayer to overlap partly with the MoS$_2$ one. (d) Optical microscope image of a device based on the structure shown in (c) with Pt/Au contacts enabling separate transport measurements to be done on the three regions (MoS$_2$ monolayer, WSe$_2$ monolayer and their interface). The scale bars in all images are 5 $ \mu $m. (e) Schematics of an ionic-liquid gated FET based on a device structure comprising two monolayers and their interface, such as the one shown in (d). The schematics also shows -- not to scale -- the gate and the reference electrode, as well as the typical bias/measurement configuration needed to perform the transistor electrical characterization.}
	\label {fig:ref 01}
	\vskip 20cm
\end{figure}

Figure 2a shows the source-drain current (\textit{$I_{SD}$}) measured on the different parts of a device based on WSe$_2$ and MoSe$_2$ monolayers, as a function of gate voltage \textit{$V_G$} (\textit{$V_{SD}$} = 50 mV; see the schematics on top). The left, center, and right panels (blue, red, and green curves) show the current flowing through the WSe$_2$ monolayer, the WSe$_2$/MoSe$_2$ interface, and the MoSe$_2$ monolayer, respectively. Excellent ambipolar characteristics are observed in all cases, demonstrating that for the monolayers, as well as for the interface, sweeping \textit{$V_G$} allows shifting the chemical potential from the conduction to valence band. As discussed earlier in multiple occasions\cite{Braga2012,Jo2014, Lezama2014, ponomarev_hole_2018}, when the current is plotted as a function of reference potential \textit{$V_{ref}$}, these measurements provide spectroscopic information. That is because, due to the very large capacitance of the ionic liquid, a change in \textit{$V_{ref}$}  corresponds to a shift in chemical potential $\Delta \mu = e V_{ref}$ as long as the density of states in the system is small (which is the case if the chemical potential is shifted inside the gap of a semiconductor)\cite{Braga2012}.\\

\begin{figure}
	\centering
	\includegraphics[width =0.9\textwidth]{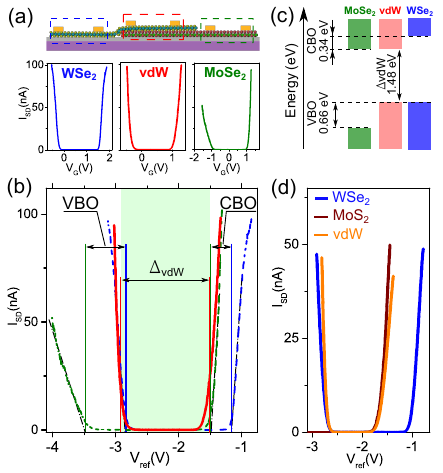}
	\caption{Spectroscopic characterization of a vdW interface and its constituents by  transport measurements in ionic liquid gated devices. (a) Top panel: Schematic representation of device configuration, showing the three different parts measured (left: WSe$_2$ monolayer;
		center: vdW interface; right: MoSe$_2$ monolayer). Bottom panels: FET transfer curves (i.e., \textit{I$_{SD}$} as a function of \textit{V$_{G}$} at fixed \textit{V$_{SD}$}) for the monolayers and the interface (see legends), all exhibiting perfectly balanced ambipolar transport. (b) Same curves shown in (a) plotted versus reference potential \textit{V$_{ref}$} (curves of a same color in (a) and (b) represent measurements done on the same part of the structure). The current through the interface (red curve) matches the current through MoSe$_2$ (green curve) when the chemical potential is in the conduction band and the current through WSe$_2$ (blue curve) when the chemical potential is in the valence band. The black dash-dotted lines represent the linear extrapolations done to determine the electron/hole threshold voltages for each curve, from which we obtain the corresponding conduction and valence band offsets (CBO and VBO, respectively), as well as the vdW interface band gap ($\Delta_{vdW}$). (c) shows the resulting band diagram corresponding to a type II (staggered) alignment\cite{sze_physics_2006} of the MoSe$_2$ and WSe$_2$ bands. (d) shows similar curves and an identical phenomenology for the WSe$_{2}$/MoS$_{2}$ system. In MoS$_2$ monolayers hole transport is suppressed by in-gap states originating from S vacancies\cite{ponomarev_hole_2018}, and yet the interface exhibits excellent hole transport, because the interfacial valence band originates from WSe$_2$.}
	\label {fig:ref 02}
	\vskip 20cm
\end{figure}

The current \textit{$I_{SD}$} measured on the different parts of the structure plotted as a function of \textit{$V_{ref}$} is shown in Figure 2b. The current through the vdW interface -- the continuous red curve --  quite precisely overlaps with the dashed green curve (i.e., the current flowing through the MoSe$_2$ monolayer) when the conduction band of the interface is populated, and with the blue dashed curve (i.e., the current flowing through the WSe$_2$ monolayer) when the valence band is populated. It follows from this observation that the interface conduction and valence bands are respectively the conduction band of the MoSe$_2$ monolayer and the band of the WSe$_2$ monolayer. This shows directly how the bands of the vdW interface -- and therefore all other low energy properties -- are fully determined by the choice of the individual constituents. Finding that the overlap is virtually perfect indicates the absence of an energetic shift between the bands of the interface and the corresponding bands of the constituent 2D materials (or of any other significant modifications of bands in the interface region), in agreement with the expected absence of any significant interfacial dipole between the layers forming the vdW interface.\\

To make these considerations systematic and quantitative, we compare the values of threshold voltage for electron and hole conduction ($V^{h}_{th}$ and $V^{e}_{th}$), for the different parts of the device. At threshold, the chemical potential is located right at the edge of the corresponding band \cite{sze_physics_2006}, i.e., at the bottom of the conduction band at threshold for electron transport and at the top of the valence band at threshold for hole transport. Therefore, since a shift in reference potential corresponds to an equal shift in chemical potential, the difference of measured threshold voltages provides a measurement of the energetic position of the bands in the different parts of the device. Specifically, the difference between  $V^{e}_{th}$ and $V^{h}_{th}$  corresponds to the band gap in the part of the device on which the measurements are done; the difference between the values of  $V^{e}_{th}$ ($V^{h}_{th}$) measured on two different parts of the devices gives the offset of their conduction (valence) bands.\\

The threshold voltages $V^{h}_{th}$ and $V^{e}_{th}$ are determined by extrapolating to zero the source-drain current $I_{SD}$ measured as a function of $V_{ref}$ (the experimental error depends on the specific device, and is typically less than 5 \% \bibnote{It is this error that is responsible for the experimental uncertainty in the position of the bands, and hence for the band offsets and the band gap extracted from the measurements. What is most strongly affected from this uncertainty are the intra-and interlayer exciton binding energies, which are are small quantities, obtained from the difference of much larger values (that is why the relative error on the extracted exciton binding energy is large).}). We  first check the values of the band gap for the MoSe$_{2}$ and WSe$_{2}$ monolayers, and find $\Delta$(MoSe$_2$) $= e ( V^{e}_{th}$(MoSe$_2$) $- V^{h}_{th}$(MoSe$_2$) $) = 2.06$ eV and $\Delta$(WSe$_2$) $= e ( V^{e}_{th}$(WSe$_2$) $- V^{h}_{th}$(WSe$_2$) $) = $ 1.73 eV, in good agreement with values reported in the literature\cite{ugeda_giant_2014,Zhang2016,Huang2016}. For the WSe$_2$/MoSe$_2$ interface the band gap is found to be $\Delta$(WSe$_2$/MoSe$_2$) $= e ( V^{e}_{th}$(WSe$_2$/MoSe$_2$) $- V^{h}_{th}$(WSe$_2$/MoSe$_2$) $) =  1.48$ eV and the same procedure to determine the conduction band offset (CBO) and valence band offset (VBO) between the MoSe$_2$ and the WSe$_2$ monolayers gives E$_{CBO}$(WSe$_2$/MoSe$_2$) = 0.34 eV and E$_{VBO}$ (WSe$_2$/MoSe$_2$) = 0.66 eV. Figure 2c summarizes these values in a single band diagram. The same analysis performed on the WSe$_{2}$/MoS$_{2}$ system (see Figure  2d) gives a value of the interface band gap of  $\Delta_{\rm WSe_2/MoS_2}$ = 1.08 eV, comparable to what found in recent scanning tunnelling spectroscopy experiments \cite{zhang_interlayer_2017}. For this interface we could only determine the conduction band offset between the two monolayers,  E$_{CBO}$(WSe$_2$/MoS$_2$)  = 0.63 eV, because -- as we recently discussed in Ref.\citenum{ponomarev_hole_2018} -- in monolayer MoS$_{2}$ the presence of defects states near the top of the valence band prevents good quality hole conduction. In this regard, it is worth emphasizing how remarkable it is that the interface exhibits ideal ambipolar behavior, even though hole transport is drastically suppressed in one of the constituent monolayers.\\

To finalize our discussion of transport, we complete the analysis of the characteristics of FETs realized on the different vdW interfaces. Representative output curves, \textit{i.e.,} the source-drain current $I_{SD}$ as a function of source-drain voltage $V_{SD}$ for different values of gate voltage $V_G$, are shown in Figure 3a for a WSe$_2$/MoS$_2$ interface (similar data is shown for WSe$_2$/MoSe$_2$ interface, in the section S2 in the Supporting Information). Except for a small hysteresis, the output curves exhibit a virtually ideal behavior: the linear regime is observed at low $V_{SD}$, followed at larger positive $V_{SD}$ by a well-defined saturation regime, and by a steep increase in $I_{SD}$ if the source-drain bias is increased further to enter the ambipolar injection regime (\textit{i.e.,} when electrons and holes are injected at opposite contacts)\cite{kang_pedagogical_2013}. We also determined the subthreshold swing by looking at the semi-logarithmic plot of the transfer curves, and found $S= 98$ mV/dec and $S= 74$ mV/dec for electron and hole transport, respectively (see Figures 3b and 3c). Both these values are very close to the ultimate room-temperature limit of 66 mV/dec\cite{sze_physics_2006}. Together with the ideal ambipolar behavior shown in Figure 2, these measurements demonstrate that the quality of FETs realized on vdW interfaces is comparable in all regards to that of FETs realized on individual TMD monolayers.\\

\begin{figure}
	\centering
	\includegraphics[width =0.9\textwidth]{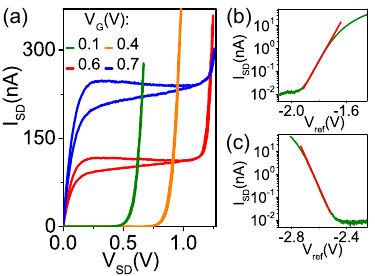}
	\caption{Electrical characteristics of a WSe$_{2}$/MoS$_{2}$ interface FET. (a) Source-drain current I$_{SD}$ as a function of source-drain bias V$_{SD}$ for different V$_{G}$ values (FET output curves) showing the behavior characteristically observed in devices based on high quality individual semiconducting monolayers. (b,c) FET transfer curves in semi-logarithmic scale in proximity of the electron (b) and hole (c) threshold. The red lines represent the linear regressions made to estimate the subthreshold swing $S$.}
	\label {fig:ref 03}
\end{figure}

Next we discuss the optoelectronic properties of the vdW interfaces, starting with PL measurements performed on the WSe$_{2}$/MoSe$_{2}$ system. The room-temperature interface PL spectrum is shown in Figure 4a (solid red curve). It consists of two main peaks centered around 1.36 eV and 1.6 eV, with the latter exhibiting a shoulder around 1.55 eV (see Figure 4b). A comparison with the spectra of the MoSe$_{2}$ (green curve in Figure 4a) and WSe$_{2}$ (blue curve in Figure 4a) monolayers shows that the 1.55 eV shoulder originates from the recombination of intralayer exctions in MoSe$_{2}$. The 1.6 eV peak is due to the recombination of charged excitons -- i.e., trions -- in WSe$_{2}$ all in agreement with values reported in the literature\cite{Mak2013,Bellus2015} . Trion formation is responsible for the red shift measured in the vdW interface as compared to the energy of the PL measured in the WSe$_2$ monolayer part of the same device, which is due to neutral excitons. In the interface region, trions form because a small amount of thermally activated charge is transferred between MoSe$_2$ and WSe$_2$; indeed the red-shift disappears at low temperature, as we show below for the case of the WSe$_2$/MoS$_2$ where the same phenomenon occurs. In short: the 1.55 eV shoulder and the 1.6 eV peak are well accounted for by intralayer transition, consistently with the fact that their energy is larger that the band gap of the WSe$_{2}$/MoSe$_{2}$. \\

\begin{figure}
	\centering
	\includegraphics[width =0.8\textwidth]{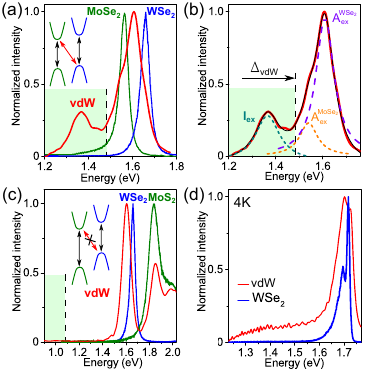}
	\caption{Photoluminescence of vdW interfaces and their constituent monolayers. (a) Normalized room-temperature PL spectra of WSe$_{2}$ (blue curve) and MoSe$_{2}$ (green curve) monolayers and of their interface (red curve). The peaks at 1.55 eV in MoSe$_{2}$ and at 1.65 eV in  WSe$_{2}$ originate from intralayer A-exciton recombination. The interface spectrum exhibits an additional peak at 1.36 eV due the recombination of interlayer excitons, formed by electrons in MoSe$_2$ and holes in WSe$_2$. The light-green shaded area represents the interval of energy inside the interface band gap $\Delta$(WSe$_2$/MoSe$_2$), extracted from transport experiments. The inset represents the alignment of the band, and the arrows indicate the optical transitions detected in the PL spectra. (b) Decomposition of WSe$_{2}$/MoSe$_{2}$ interface PL spectrum seen in (a) as a sum of the three identified radiative transitions (the formation of trions in the interface region is responsible for the red-shift of the PL peak due WSe$_2$ intralayer exciton recombination; see main text). (c) Same as (a) for the WSe$_{2}$/MoS$_2$ system. The blue, green and red curves represent the PL of the individual WSe$_{2}$ and MoS$_{2}$ monolayers, and of the interface, respectively. No sign of a radiative interlayer transition is seen in this case. Also here, the WSe$_2$ PL peak is red-shifted in the interface region due to the formation of trions, but at $T=$ 4.2 K the red-shift disappears, as shown by the PL spectra plotted in (d).}
\end{figure}

Contrary to these features, the 1.36 eV peak in the interface PL spectrum shown in Figure 4a is obviously absent in the spectra of either constituent monolayer. This peak is a manifestation of interlayer excitons formed by electrons and holes located in different layers -- a hole in WSe$_2$ and an electron in MoSe$_2$ in the present case -- with the corresponding transition represented by the diagonal red arrow in the inset of Figure 4a\cite{rivera_observation_2015,rivera_valley-polarized_2016,ross_interlayer_2017}. Consistently with this attribution, the energy of the transition is smaller than the single particle interface band gap, $\Delta$(WSe$_2$/MoSe$_2$) = 1.48 eV, as extracted from transport. With the experiments giving us both the interfacial band gap and the interfacial exciton recombination energy $E_{PL}$, the binding energy of the interfacial exciton (${\rm X^0_i}$) can be determined to be $E_{\rm X^0_i}$(WSe$_2$/MoSe$_2$) $ = \Delta$(WSe$_2$/MoSe$_2$)$- E_{\rm PL}= 120$ meV. Despite any possible reduction due to the enhanced screening caused by the presence of the ionic liquid, this is a rather large value, as expected from recent estimates\cite{Wilson2017}. Importantly, finding that the interfacial exciton (${\rm X^0_i}$) decays through a radiative transition strongly supports the conclusion that the band gap of the WSe$_{2}$/MoSe$_{2}$ interface is direct in $k-$space\cite{yu_anomalous_2015} . That is because the constituent monolayers are nearly perfectly lattice matched and their crystallographic orientations have been intentionally aligned during the device fabrication. \\

Following the same logic\cite{yu_anomalous_2015} , no PL from interlayer excitons is expected in WSe$_{2}$/MoS$_{2}$ interfaces irrespective of the alignment of the crystallographic axis of the constituent monolayers, because of a 5 \% mismatch in their lattice constants. The mismatch implies that the K/K' points of the WSe$_2$ and MoS$_2$ monolayers are located at different points in $k-$space, preventing direct interlayer transitions\cite{Zhu2017}. The PL spectra for the WSe$_{2}$/MoS$_{2}$ system is shown in Figure 4c, for a device in which the crystallographic axis of the monolayers have been carefully aligned. The PL from the interface (red line) exhibits two peaks, at 1.59 eV and at 1.84 eV. The latter coincides with the peak originating from direct intralayer exciton transitions in MoS$_2$, as it can be inferred from the PL spectrum measured on the MoS$_2$ monolayer (green line in Figure 4c). Similarly to the case of the WSe$_2$/MoSe$_2$ interface, the 1.59 eV peak is due to intralayer trion recombination in WSe$_2$, and is red-shifted relative to the recombination of neutral excitons in the isolated WSe$_2$ (blue line in Figure  4c). Here as well, trions in WSe$_2$ are formed in the interface region due to a small density of thermally activated charge carriers transferred from MoS$_2$. Indeed, when the PL spectrum of the interface and of WSe$_2$ are measured at $T=4.2$ K -- with thermal transfer of carriers fully suppressed -- the shift disappears and the position of the peak measured in the isolated WSe$_2$ monolayer coincides with the position of the peak measured in the WSe$_2$ forming the interface (see Figure 4d; the only difference is that the peak of the isolated monolayer is sharper).\\

The most relevant aspect of these measurements is the absence of any features at energy smaller than those originating from intralayer transitions in the individual monolayers. As the single particle band gap extracted earlier from transport experiments is $\Delta {\rm (WSe_2/MoS_2)}$ = 1.08 eV, any possible interlayer transition should be visible below this energy. However, no signal is observed here even though our spectrometer is sensitive down to 0.8 eV. Several devices with different rotational alignment were studied with no significant difference in their PL response: in no case a peak with energy smaller than 1.08 eV was observed. The experimental results therefore fully support the conclusion that the band gap of the WSe$_{2}$/MoS$_{2}$ interface is indirect in $k-$space. \\

More information about the interface interband transitions can be obtained from photocurrent spectroscopy\cite{ross_interlayer_2017,lee_atomically_2014}. Whereas the outcome of PL experiments strongly depends on the competition between radiative and non-radiative decay processes, PC spectroscopy reveals details of optical interband transitions in a way similar to optical absorption measurements\cite{Miller1985,collins_photocurrent_1986}. Figure 5a,b show the short-circuit current \textit{\textit{I$_{SC}$}} --$\textit{i.e.}$, the PC measured with short-circuited contacts -- for the WSe$_2$/MoSe$_2$ and the WSe$_2$/MoS$_2$ interfaces, as function of laser excitation energy. We discuss exclusively the spectral dependence of the PC, and not on its absolute magnitude, which critically depends on details of the experiments (the device geometry, the applied gate voltage, the spatial profile of the incident light, etc.; See section S4 in the Supporting Information).\\

\begin{figure}
	\centering
	\includegraphics[width =0.6\textwidth]{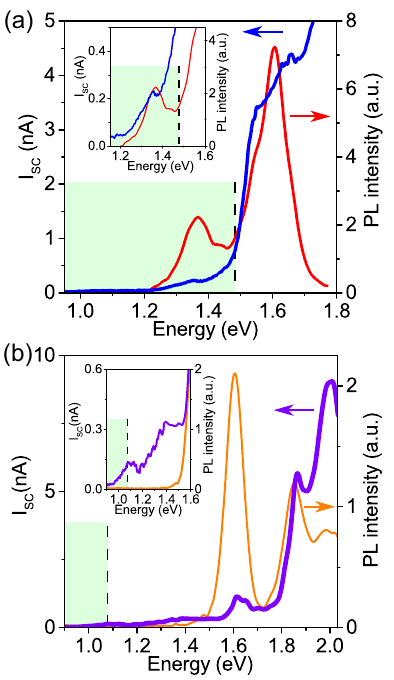}
	\caption{Photocurrent spectroscopy. (a) The short-circuit current, \textit{I$_{SC}$}, measured on a WSe$_2$/MoSe$_2$ interface as function of incident photon energy  (blue line) is compared to the interface PL spectrum (red line). The light-green shaded region marks the energy interval below the interface band gap, as extracted from transport measurements.  The inset zooms-in on the two curves at low energy. It is apparent that the PC onset occurs in  correspondence of the interlayer exciton transition and that a steep increase in \textit{I$_{SC}$} occurs just below 1.5 eV, \textit{i.e.,} in correspondence of interfacial interband transitions (the interface band gap is $\Delta$(WSe$_2$/MoSe$_2$)=1.48 eV). (b) PC spectrum measured on the WSe$_2$/MoS$_2$ interface (blue curve) together with the corresponding PL spectrum (orange curve). The inset zooms-in on the photocurrent at energy smaller than the intralayer transitions, showing that a slowly increasing photocurrent is present, starting from approximately 1 eV, close to the interface band gap $\Delta$(WSe$_2$/MoS$_2$)=1.08 eV. The slow increase is consistent with the indirect nature of the interlayer transitions.}
	\label {fig:ref 05}
\end{figure}

As it is apparent from a direct comparison with the PL spectra (see Figure  5a,b and their insets), both interfaces exhibit a measurable PC starting at energies well below the energy of the excitonic transition in the constituent monolayers (see also the comparison of the PC measured on each interface and on the constituent monolayers, Figure S2 of the Supporting Information). For the WSe$_2$/MoSe$_2$ interface, the PC becomes measurable near the onset of the PL line associated to the interlayer exciton radiative decay (exhibiting a shallow maximum near the PL peak, see the inset of Figure  5a). It then starts to increase steeply in correspondence of the vdW interface gap $\Delta$(WSe$_2$/MoSe$_2$) = 1.48 eV. This behavior is expected, since interlayer transitions in the WSe$_2$/MoSe$_2$ interface are direct in $k-$space. For the WSe$_2$/MoS$_2$ interface the PC onset is just under 1.08 eV (Figure 5b and its inset), and the PC  increases only slowly as the energy is increased up to approximately 1.6 eV, at which point a steeper enhancement is observed due to direct intralayer exciton absorption in WSe$_2$. Finding that the PC onset energy matches the value of band gap is interesting, since the PL of the  WSe$_2$/MoS$_2$ interface  did not show any feature at $\Delta$(WSe$_2$/MoS$_2$) = 1.08 eV. Overall the expected behavior, and in particular the slow increase in PC at energies smaller than the intralayer transitions, is consistent with the interfacial band gap of WSe$_2$/MoS$_2$ being indirect in $k-$space.  For both interfaces, therefore, the PC measurements fully support the conclusions drawn from transport and from PL measurements.\\

In summary, our work demonstrates an approach to characterize systematically the basic properties of semiconducting van der Waals interfaces, enabling a quantitative relation with the corresponding properties of the constituent 2D materials to be established. More specifically, our experiments show directly how the interfacial conduction and valence bands are independently determined by the conduction and valence bands of the constituent monolayers, they allow the full quantitative determination of the energetics of the system (including the single-particle band gaps of the two materials and of the interface, the values of the band offsets, and the exciton binding energies), and provide definite information about the direct/indirect nature of inter-band transitions. These conclusions have been obtained for two specific interfaces, but the experimental method that we have demonstrated can be applied to any other system of the type discussed here. Even more important in the broader context, the experiments show that the opto-electronic response of  semiconducting vdW interfaces  -- including transport in transistor devices, photoluminescence, photocurrent, etc. -- is virtually indistinguishable from that of an individual semiconducting monolayer. Nevertheless, there is clearly a very important difference between the two cases, namely that the material properties of an individual monolayer are given and cannot be modified, whereas the properties of interfaces can be deterministically defined by appropriately selecting the constituent monolayers in the vast portfolio of existing semiconducting 2D materials. As such, vdW interfaces are composite systems that behave in all regards as artificial semiconductors with properties that can be engineered by design at the assembly stage.

\begin{acknowledgement}
We gratefully acknowledge A. Ferreira for continuous technical support and Hugo Henck for his assistance with the preparation of the figures. A.F.M. gratefully acknowledge financial support from the Swiss National Science Foundation and from the EU Graphene Flagship project. N.U. acknowledges funding from an Ambizione grant of the Swiss National Science Foundation.
\end{acknowledgement}

\section*{Competing financial interests}
The authors declare no competing financial interests.

\begin{suppinfo}
	Methods, output characteristics of WSe$_2$/MoSe$_2$ interface, photocurrent spectra and scanning photocurrent microscopy
\end{suppinfo}

\providecommand{\latin}[1]{#1}
\makeatletter
\providecommand{\doi}
  {\begingroup\let\do\@makeother\dospecials
  \catcode`\{=1 \catcode`\}=2 \doi@aux}
\providecommand{\doi@aux}[1]{\endgroup\texttt{#1}}
\makeatother
\providecommand*\mcitethebibliography{\thebibliography}
\csname @ifundefined\endcsname{endmcitethebibliography}
  {\let\endmcitethebibliography\endthebibliography}{}

%\vspace{6\baselineskip}
%\begin{figure}
%	\centering
%	\includegraphics[width =\textwidth]{TOC}
%
%	\label {TOC}
%\end{figure}


\begin{mcitethebibliography}{49}
\providecommand*\natexlab[1]{#1}
\providecommand*\mciteSetBstSublistMode[1]{}
\providecommand*\mciteSetBstMaxWidthForm[2]{}
\providecommand*\mciteBstWouldAddEndPuncttrue
  {\def\EndOfBibitem{\unskip.}}
\providecommand*\mciteBstWouldAddEndPunctfalse
  {\let\EndOfBibitem\relax}
\providecommand*\mciteSetBstMidEndSepPunct[3]{}
\providecommand*\mciteSetBstSublistLabelBeginEnd[3]{}
\providecommand*\EndOfBibitem{}
\mciteSetBstSublistMode{f}
\mciteSetBstMaxWidthForm{subitem}{(\alph{mcitesubitemcount})}
\mciteSetBstSublistLabelBeginEnd
  {\mcitemaxwidthsubitemform\space}
  {\relax}
  {\relax}

\bibitem[Geim and Grigorieva(2013)Geim, and Grigorieva]{geim_van_2013}
Geim,~A.~K.; Grigorieva,~I.~V. Van der Waals heterostructures.
  \emph{Nature} \textbf{2013}, \emph{499}, 419--425\relax
\mciteBstWouldAddEndPuncttrue
\mciteSetBstMidEndSepPunct{\mcitedefaultmidpunct}
{\mcitedefaultendpunct}{\mcitedefaultseppunct}\relax
\EndOfBibitem
\bibitem[Novoselov \latin{et~al.}(2016)Novoselov, Mishchenko, Carvalho, and
  Neto]{novoselov_2d_2016}
Novoselov,~K.~S.; Mishchenko,~A.; Carvalho,~A.; Neto,~A. H.~C. 2D materials and
  van der {Waals} heterostructures. \emph{Science} \textbf{2016}, \emph{353},
  aac9439\relax
\mciteBstWouldAddEndPuncttrue
\mciteSetBstMidEndSepPunct{\mcitedefaultmidpunct}
{\mcitedefaultendpunct}{\mcitedefaultseppunct}\relax
\EndOfBibitem
\bibitem[Iannaccone \latin{et~al.}(2018)Iannaccone, Bonaccorso, Colombo, and
  Fiori]{iannaccone_quantum_2018}
Iannaccone,~G.; Bonaccorso,~F.; Colombo,~L.; Fiori,~G. Quantum Engineering of
  Transistors Based on 2D materials Heterostructures. \emph{Nature
  Nanotechnology} \textbf{2018}, \emph{13}, 183--191\relax
\mciteBstWouldAddEndPuncttrue
\mciteSetBstMidEndSepPunct{\mcitedefaultmidpunct}
{\mcitedefaultendpunct}{\mcitedefaultseppunct}\relax
\EndOfBibitem
\bibitem[Splendiani \latin{et~al.}(2010)Splendiani, Sun, Zhang, Li, Kim, Chim,
  Galli, and Wang]{splendiani_emerging_2010}
Splendiani,~A.; Sun,~L.; Zhang,~Y.; Li,~T.; Kim,~J.; Chim,~C.-Y.; Galli,~G.;
  Wang,~F. Emerging Photoluminescence in Monolayer MoS$_{2}$.
  \emph{Nano Lett.} \textbf{2010}, \emph{10}, 1271--1275\relax
\mciteBstWouldAddEndPuncttrue
\mciteSetBstMidEndSepPunct{\mcitedefaultmidpunct}
{\mcitedefaultendpunct}{\mcitedefaultseppunct}\relax
\EndOfBibitem
\bibitem[Mak \latin{et~al.}(2010)Mak, Lee, Hone, Shan, and
  Heinz]{mak_atomically_2010}
Mak,~K.~F.; Lee,~C.; Hone,~J.; Shan,~J.; Heinz,~T.~F. Atomically Thin
  MoS$_{2}$: A New Direct-Gap Semiconductor. \emph{Phys. Rev. Lett.} \textbf{2010}, \emph{105},
  136805\relax
\mciteBstWouldAddEndPuncttrue
\mciteSetBstMidEndSepPunct{\mcitedefaultmidpunct}
{\mcitedefaultendpunct}{\mcitedefaultseppunct}\relax
\EndOfBibitem
\bibitem[Wang \latin{et~al.}(2015)Wang, Gutiérrez-Lezama, Barreteau, Ubrig,
  Giannini, and Morpurgo]{wang_tuning_2015}
Wang,~L.; Gutiérrez-Lezama,~I.; Barreteau,~C.; Ubrig,~N.; Giannini,~E.;
  Morpurgo,~A.~F. Tuning Magnetotransport in a Compensated Semimetal at the
  Atomic Scale. \emph{Nat. Commun.} \textbf{2015}, \emph{6}, 8892\relax
\mciteBstWouldAddEndPuncttrue
\mciteSetBstMidEndSepPunct{\mcitedefaultmidpunct}
{\mcitedefaultendpunct}{\mcitedefaultseppunct}\relax
\EndOfBibitem
\bibitem[Fei \latin{et~al.}(2017)Fei, Palomaki, Wu, Zhao, Cai, Sun, Nguyen,
  Finney, Xu, and Cobden]{fei_edge_2017}
Fei,~Z.; Palomaki,~T.; Wu,~S.; Zhao,~W.; Cai,~X.; Sun,~B.; Nguyen,~P.;
  Finney,~J.; Xu,~X.; Cobden,~D.~H. Edge Conduction in Monolayer
  WTe$_{2}$. \emph{Nat. Phys.}
  \textbf{2017}, \emph{13}, 677--682\relax
\mciteBstWouldAddEndPuncttrue
\mciteSetBstMidEndSepPunct{\mcitedefaultmidpunct}
{\mcitedefaultendpunct}{\mcitedefaultseppunct}\relax
\EndOfBibitem
\bibitem[Gong \latin{et~al.}(2017)Gong, Li, Li, Ji, Stern, Xia, Cao, Bao, Wang,
  Wang, Qiu, Cava, Louie, Xia, and Zhang]{Gong2017}
Gong,~C.; Li,~L.; Li,~Z.; Ji,~H.; Stern,~A.; Xia,~Y.; Cao,~T.; Bao,~W.;
  Wang,~C.; Wang,~Y.; Qiu,~Z. Q.; Cava,~R.~J.; Louie,~S.G.; Xia,~J.; Zhang,~X. Discovery of Intrinsic Ferromagnetism in Two-Dimensional van der Waals Crystals. \emph{Nature} \textbf{2017},
  \emph{546}, 265--269\relax
\mciteBstWouldAddEndPuncttrue
\mciteSetBstMidEndSepPunct{\mcitedefaultmidpunct}
{\mcitedefaultendpunct}{\mcitedefaultseppunct}\relax
\EndOfBibitem
\bibitem[Huang \latin{et~al.}(2017)Huang, Clark, Navarro-Moratalla, Klein,
  Cheng, Seyler, Zhong, Schmidgall, McGuire, Cobden, Yao, Xiao,
  Jarillo-Herrero, and Xu]{huang_layer-dependent_2017}
Huang,~B.; Clark,~G.; Navarro-Moratalla,~E.; Klein,~D.~R.; Cheng,~R.;
  Seyler,~K.~L.; Zhong,~D.; Schmidgall,~E.; McGuire,~M.~A.; Cobden,~D.~H.; Yao,~W.; Xiao,~D.; Jarillo-Herrero,~P.; Xu,~X. Layer-Dependent Ferromagnetism in a van der Waals Crystal Down to the Monolayer Limit. \emph{Nature} \textbf{2017}, \emph{546},
  270--273\relax
\mciteBstWouldAddEndPuncttrue
\mciteSetBstMidEndSepPunct{\mcitedefaultmidpunct}
{\mcitedefaultendpunct}{\mcitedefaultseppunct}\relax
\EndOfBibitem
\bibitem[Cao \latin{et~al.}(2015)Cao, Mishchenko, Yu, Khestanova, Rooney,
  Prestat, Kretinin, Blake, Shalom, Woods, Chapman, Balakrishnan, Grigorieva,
  Novoselov, Piot, Potemski, Watanabe, Taniguchi, Haigh, Geim, and
  Gorbachev]{Cao2015}
Cao,~Y.; Mishchenko,~A.; Yu,~G.~L.; Khestanova,~E.; Rooney,~A.~P.; Prestat,~E.;
  Kretinin,~A.~V.; Blake,~P.; Shalom,~M.~B.; Woods,~C.; Chapman,~J.; Balakrishnan,~G.; Grigorieva,~ I. V.; Novoselov,~K.S.; Piot,~B.A.; Potemski,~M.; Watanabe,~K.; Taniguchi,~T.; Haigh,~S. J.; Geim,~A. K.; Gorbachev,~R.V. Quality Heterostructures from Two-Dimensional Crystals Unstable in Air by Their Assembly in Inert Atmosphere. \emph{Nano Lett.} \textbf{2015},
  \emph{15}, 4914--4921\relax
\mciteBstWouldAddEndPuncttrue
\mciteSetBstMidEndSepPunct{\mcitedefaultmidpunct}
{\mcitedefaultendpunct}{\mcitedefaultseppunct}\relax
\EndOfBibitem
\bibitem[Xi \latin{et~al.}(2016)Xi, Wang, Zhao, Park, Law, Berger, Forró,
  Shan, and Mak]{xi_ising_2016}
Xi,~X.; Wang,~Z.; Zhao,~W.; Park,~J.-H.; Law,~K.~T.; Berger,~H.; Forró,~L.;
  Shan,~J.; Mak,~K.~F. Ising Pairing in Superconducting
  NbSe$_{2}$ Atomic Layers. \emph{Nat. Phys.} \textbf{2016}, \emph{12}, 139--143\relax
\mciteBstWouldAddEndPuncttrue
\mciteSetBstMidEndSepPunct{\mcitedefaultmidpunct}
{\mcitedefaultendpunct}{\mcitedefaultseppunct}\relax
\EndOfBibitem
\bibitem[Yankowitz \latin{et~al.}(2012)Yankowitz, Xue, Cormode,
  Sanchez-Yamagishi, Watanabe, Taniguchi, Jarillo-Herrero, Jacquod, and
  LeRoy]{yankowitz_emergence_2012}
Yankowitz,~M.; Xue,~J.; Cormode,~D.; Sanchez-Yamagishi,~J.~D.; Watanabe,~K.;
  Taniguchi,~T.; Jarillo-Herrero,~P.; Jacquod,~P.; LeRoy,~B.~J. Emergence of
  Superlattice Dirac Points in Graphene on Hexagonal Boron Nitride.
  \emph{Nat. Phys.} \textbf{2012}, \emph{8}, 382--386\relax
\mciteBstWouldAddEndPuncttrue
\mciteSetBstMidEndSepPunct{\mcitedefaultmidpunct}
{\mcitedefaultendpunct}{\mcitedefaultseppunct}\relax
\EndOfBibitem
\bibitem[Dean \latin{et~al.}(2013)Dean, Wang, Maher, Forsythe, Ghahari, Gao,
  Katoch, Ishigami, Moon, Koshino, Taniguchi, Watanabe, Shepard, Hone, and
  Kim]{dean_hofstadters_2013}
Dean,~C.~R.; Wang,~L.; Maher,~P.; Forsythe,~C.; Ghahari,~F.; Gao,~Y.;
  Katoch,~J.; Ishigami,~M.; Moon,~P.; Koshino,~M.; Taniguchi,~T.; Watanabe,~K.; Shepard,~KL.; Hone,~J.; Kim,~P. Hofstadter's
  Butterfly and the Fractal Quantum Hall Effect in Moiré Superlattices.
  \emph{Nature} \textbf{2013}, \emph{497}, 598--602\relax
\mciteBstWouldAddEndPuncttrue
\mciteSetBstMidEndSepPunct{\mcitedefaultmidpunct}
{\mcitedefaultendpunct}{\mcitedefaultseppunct}\relax
\EndOfBibitem
\bibitem[Ponomarenko \latin{et~al.}(2013)Ponomarenko, Gorbachev, Yu, Elias,
  Jalil, Patel, Mishchenko, Mayorov, Woods, Wallbank, Mucha-Kruczynski, Piot,
  Potemski, Grigorieva, Novoselov, Guinea, Fal'ko, and
  Geim]{ponomarenko_cloning_2013}
Ponomarenko,~L.~A.; Gorbachev,~R.~V.; Yu,~G.~L.; Elias,~D.~C.; Jalil,~R.;
  Patel,~A.~A.; Mishchenko,~A.; Mayorov,~A.~S.; Woods,~C.~R.; Wallbank,~J.~R.; Mucha-Kruczynski,~M.; Piot~B.A.; Potemski~M.; Grigorieva~I.V.; Novoselov~K.S.; Guinea~F.; Fal'ko~V.I.; Geim,~A.K. Cloning of Dirac Fermions in Graphene Superlattices.
  \emph{Nature} \textbf{2013}, \emph{497}, 594--597\relax
\mciteBstWouldAddEndPuncttrue
\mciteSetBstMidEndSepPunct{\mcitedefaultmidpunct}
{\mcitedefaultendpunct}{\mcitedefaultseppunct}\relax
\EndOfBibitem
\bibitem[Wang \latin{et~al.}(2015)Wang, Ki, Chen, Berger, MacDonald, and
  Morpurgo]{wang_strong_2015-1}
Wang,~Z.; Ki,~D.-K.; Chen,~H.; Berger,~H.; MacDonald,~A.~H.; Morpurgo,~A.~F.
  Strong Interface-Induced Spin–Orbit Interaction in Graphene on
  WS$_{2}$. \emph{Nat. Commun.}
  \textbf{2015}, \emph{6}, 8339\relax
\mciteBstWouldAddEndPuncttrue
\mciteSetBstMidEndSepPunct{\mcitedefaultmidpunct}
{\mcitedefaultendpunct}{\mcitedefaultseppunct}\relax
\EndOfBibitem
\bibitem[Wang \latin{et~al.}(2016)Wang, Ki, Khoo, Mauro, Berger, Levitov, and
  Morpurgo]{Wang2016}
Wang,~Z.; Ki,~D.~K.; Khoo,~J.~Y.; Mauro,~D.; Berger,~H.; Levitov,~L.~S.;
  Morpurgo,~A.~F. Origin and Magnitude of 'Designer' Spin-Orbit Interaction in
  Graphene on Semiconducting Transition Metal Dichalcogenides. \emph{Phys. Rev.
  X} \textbf{2016}, \emph{6}, 1--15\relax
\mciteBstWouldAddEndPuncttrue
\mciteSetBstMidEndSepPunct{\mcitedefaultmidpunct}
{\mcitedefaultendpunct}{\mcitedefaultseppunct}\relax
\EndOfBibitem
\bibitem[Cao \latin{et~al.}(2018)Cao, Fatemi, Fang, Watanabe, Taniguchi,
  Kaxiras, and Jarillo-Herrero]{cao_unconventional_2018}
Cao,~Y.; Fatemi,~V.; Fang,~S.; Watanabe,~K.; Taniguchi,~T.; Kaxiras,~E.;
  Jarillo-Herrero,~P. Unconventional Superconductivity in Magic-Angle Graphene
  Superlattices. \emph{Nature} \textbf{2018}, \emph{556}, 43--50\relax
\mciteBstWouldAddEndPuncttrue
\mciteSetBstMidEndSepPunct{\mcitedefaultmidpunct}
{\mcitedefaultendpunct}{\mcitedefaultseppunct}\relax
\EndOfBibitem
\bibitem[Not()]{Note-1}
The case of a band alignment resulting in either conduction or the valence
  bands of the two monolayers that are degenerate in energy is not considered
  here. Despite providing even more possibilities to control the interfacial
  properties, that case is more complex because the hybridization of the states
  in the degenerate bands of the two materials can not be neglected.\relax
\mciteBstWouldAddEndPunctfalse
\mciteSetBstMidEndSepPunct{\mcitedefaultmidpunct}
{}{\mcitedefaultseppunct}\relax
\EndOfBibitem
\bibitem[Hong \latin{et~al.}(2014)Hong, Kim, Shi, Zhang, Jin, Sun, Tongay, Wu,
  Zhang, and Wang]{hong_ultrafast_2014}
Hong,~X.; Kim,~J.; Shi,~S.-F.; Zhang,~Y.; Jin,~C.; Sun,~Y.; Tongay,~S.; Wu,~J.;
  Zhang,~Y.; Wang,~F. Ultrafast Charge Transfer in Atomically Thin
  {MoS}$_{\textrm{2}}$/{WS}$_{\textrm{2}}$ heterostructures. \emph{Nature
  Nanotechnology} \textbf{2014}, \emph{9}, 682--686\relax
\mciteBstWouldAddEndPuncttrue
\mciteSetBstMidEndSepPunct{\mcitedefaultmidpunct}
{\mcitedefaultendpunct}{\mcitedefaultseppunct}\relax
\EndOfBibitem
\bibitem[Rivera \latin{et~al.}(2015)Rivera, Schaibley, Jones, Ross, Wu,
  Aivazian, Klement, Seyler, Clark, Ghimire, Yan, Mandrus, Yao, and
  Xu]{rivera_observation_2015}
Rivera,~P.; Schaibley,~J.~R.; Jones,~A.~M.; Ross,~J.~S.; Wu,~S.; Aivazian,~G.;
  Klement,~P.; Seyler,~K.; Clark,~G.; Ghimire,~N.~J.; Yan,~J.; Mandrus,~D.G.; Yao,~W.; Xu,~X.
  Observation of Long-Lived Interlayer Excitons in Monolayer
  MoSe$_{2}$-WSe$_{2}$ Heterostructures. \emph{Nat. Commun.} \textbf{2015},
  \emph{6}, 6242\relax
\mciteBstWouldAddEndPuncttrue
\mciteSetBstMidEndSepPunct{\mcitedefaultmidpunct}
{\mcitedefaultendpunct}{\mcitedefaultseppunct}\relax
\EndOfBibitem
\bibitem[Wang \latin{et~al.}(2016)Wang, Huang, Tian, Ceballos, Lin,
  Mahjouri-Samani, Boulesbaa, Puretzky, Rouleau, Yoon, Zhao, Xiao, Duscher, and
  Geohegan]{wang_interlayer_2016}
Wang,~K.; Huang,~B.; Tian,~M.; Ceballos,~F.; Lin,~M.-W.; Mahjouri-Samani,~M.;
  Boulesbaa,~A.; Puretzky,~A.~A.; Rouleau,~C.~M.; Yoon,~M.; Zhao,~H.; Xiao,~K.; Duscher,~G.; Geohegan,~D.B. Interlayer Coupling in Twisted WSe$_{2}$/WS$_{2}$ Bilayer Heterostructures
  Revealed by Optical Spectroscopy. \emph{ACS Nano} \textbf{2016},
  \emph{10}, 6612--6622\relax
\mciteBstWouldAddEndPuncttrue
\mciteSetBstMidEndSepPunct{\mcitedefaultmidpunct}
{\mcitedefaultendpunct}{\mcitedefaultseppunct}\relax
\EndOfBibitem
\bibitem[Kim \latin{et~al.}(2017)Kim, Jin, Chen, Cai, Zhao, Lee, Kahn,
  Watanabe, Taniguchi, Tongay, Crommie, and Wang]{kim_observation_2017}
Kim,~J.; Jin,~C.; Chen,~B.; Cai,~H.; Zhao,~T.; Lee,~P.; Kahn,~S.; Watanabe,~K.;
  Taniguchi,~T.; Tongay,~S.; Crommie,~M.F.; Wang,~F. Observation of Ultralong Valley
  Lifetime in WSe$_{2}$/MoS$_{2}$ Heterostructures. \emph{Sci. Adv.}
  \textbf{2017}, \emph{3}, e1700518\relax
\mciteBstWouldAddEndPuncttrue
\mciteSetBstMidEndSepPunct{\mcitedefaultmidpunct}
{\mcitedefaultendpunct}{\mcitedefaultseppunct}\relax
\EndOfBibitem
\bibitem[Panzer \latin{et~al.}(2005)Panzer, Newman, and Frisbie]{Panzer2005}
Panzer,~M.~J.; Newman,~C.~R.; Frisbie,~C.~D. Low-Voltage Operation of a
  Pentacene Field-Effect Transistor with a Polymer Electrolyte Gate Dielectric.
  \emph{Appl. Phys. Lett.} \textbf{2005}, \emph{86}, 1--3\relax
\mciteBstWouldAddEndPuncttrue
\mciteSetBstMidEndSepPunct{\mcitedefaultmidpunct}
{\mcitedefaultendpunct}{\mcitedefaultseppunct}\relax
\EndOfBibitem
\bibitem[Shimotani \latin{et~al.}(2006)Shimotani, Asanuma, Takeya, and
  Iwasa]{Shimotani2006}
Shimotani,~H.; Asanuma,~H.; Takeya,~J.; Iwasa,~Y. Electrolyte-Gated Charge
  Accumulation in Organic Single Crystals. \emph{Appl. Phys. Lett.}
  \textbf{2006}, \emph{89}\relax
\mciteBstWouldAddEndPuncttrue
\mciteSetBstMidEndSepPunct{\mcitedefaultmidpunct}
{\mcitedefaultendpunct}{\mcitedefaultseppunct}\relax
\EndOfBibitem
\bibitem[Misra \latin{et~al.}(2007)Misra, McCarthy, and Hebard]{Misra2007}
Misra,~R.; McCarthy,~M.; Hebard,~A.~F. Electric Field Gating with Ionic
  Liquids. \emph{Appl. Phys. Lett.} \textbf{2007}, \emph{90}, 2005--2008\relax
\mciteBstWouldAddEndPuncttrue
\mciteSetBstMidEndSepPunct{\mcitedefaultmidpunct}
{\mcitedefaultendpunct}{\mcitedefaultseppunct}\relax
\EndOfBibitem
\bibitem[Xie and Frisbie()Xie, and Frisbie]{xie_organic_2011}
Xie,~W.; Frisbie,~C.~D. Organic Electrical Double Layer Transistors Based on Rubrene Single Crystals: Examining Transport at High Surface Charge Densities above 10$^{13}$ cm$^{-2}$ \emph{J. Phys. Chem. C} \textbf{2011}, 115 (29) 14360--14368\relax 
\mciteBstWouldAddEndPuncttrue
\mciteSetBstMidEndSepPunct{\mcitedefaultmidpunct}
{\mcitedefaultendpunct}{\mcitedefaultseppunct}\relax
\EndOfBibitem
\bibitem[Braga \latin{et~al.}(2012)Braga, Gutiérrez~Lezama, Berger, and
  Morpurgo]{Braga2012}
Braga,~D.; Gutiérrez~Lezama,~I.; Berger,~H.; Morpurgo,~A.~F. Quantitative
  Determination of the Band Gap of WS$_{2}$ with Ambipolar Ionic Liquid-Gated Transistors. \emph{Nano Lett.} \textbf{2012}, \emph{12}, 5218--5223\relax
\mciteBstWouldAddEndPuncttrue
\mciteSetBstMidEndSepPunct{\mcitedefaultmidpunct}
{\mcitedefaultendpunct}{\mcitedefaultseppunct}\relax
\EndOfBibitem
\bibitem[Jo \latin{et~al.}(2014)Jo, Ubrig, Berger, Kuzmenko, and
  Morpurgo]{Jo2014}
Jo,~S.; Ubrig,~N.; Berger,~H.; Kuzmenko,~A.~B.; Morpurgo,~A.~F. Mono- and
  Bilayer WS$_{2}$ Light-Emitting Transistors. \emph{Nano Lett.} \textbf{2014}, \emph{14(4)}, 2019--2025.\relax
\mciteBstWouldAddEndPunctfalse
\mciteSetBstMidEndSepPunct{\mcitedefaultmidpunct}
{}{\mcitedefaultseppunct}\relax
\EndOfBibitem
\bibitem[Lezama \latin{et~al.}(2014)Lezama, Ubaldini, Longobardi, Giannini,
  Renner, Kuzmenko, and Morpurgo]{Lezama2014}
Lezama,~I.~G.; Ubaldini,~A.; Longobardi,~M.; Giannini,~E.; Renner,~C.;
  Kuzmenko,~A.~B.; Morpurgo,~A.~F. Surface Transport and Band Gap Structure of
  Exfoliated 2H-MoTe$_{2}$ crystals. \emph{2D Mater.} \textbf{2014}, \emph{1}, 021002\relax
\mciteBstWouldAddEndPuncttrue
\mciteSetBstMidEndSepPunct{\mcitedefaultmidpunct}
{\mcitedefaultendpunct}{\mcitedefaultseppunct}\relax
\EndOfBibitem
\bibitem[Ponomarev \latin{et~al.}(2018)Ponomarev, Pasztor, Waelchli, Scarfato,
  Ubrig, Renner, and Morpurgo]{ponomarev_hole_2018}
Ponomarev,~E.; Pasztor,~A.; Waelchli,~A.; Scarfato,~A.; Ubrig,~N.; Renner,~C.;
  Morpurgo,~A.~F. Hole Transport in Exfoliated Monolayer MoS$_{2}$. \emph{ACS Nano} \textbf{2018}, \emph{12}, 2669--2676\relax
\mciteBstWouldAddEndPuncttrue
\mciteSetBstMidEndSepPunct{\mcitedefaultmidpunct}
{\mcitedefaultendpunct}{\mcitedefaultseppunct}\relax
\EndOfBibitem
\bibitem[Yu \latin{et~al.}(2015)Yu, Wang, Tong, Xu, and Yao]{yu_anomalous_2015}
Yu,~H.; Wang,~Y.; Tong,~Q.; Xu,~X.; Yao,~W. Anomalous Light Cones and Valley Optical Selection Rules of InterlayerExcitons in Twisted Heterobilayers.
  \emph{Phys. Rev. Lett.} \textbf{2015}, \emph{115}, 1--5\relax
\mciteBstWouldAddEndPuncttrue
\mciteSetBstMidEndSepPunct{\mcitedefaultmidpunct}
{\mcitedefaultendpunct}{\mcitedefaultseppunct}\relax
\EndOfBibitem
\bibitem[Zomer \latin{et~al.}(2011)Zomer, Dash, Tombros, and van
  Wees]{zomer_transfer_2011}
Zomer,~P.~J.; Dash,~S.~P.; Tombros,~N.; van Wees,~B.~J. A Transfer Technique
  for High Mobility Graphene Devices on Commercially Available Hexagonal Boron
  Nitride. \emph{Appl. Phys. Lett.} \textbf{2011}, \emph{99}, 232104\relax
\mciteBstWouldAddEndPuncttrue
\mciteSetBstMidEndSepPunct{\mcitedefaultmidpunct}
{\mcitedefaultendpunct}{\mcitedefaultseppunct}\relax
\EndOfBibitem
\bibitem[Sze and Ng(2006)Sze, and Ng]{sze_physics_2006}
Sze,~S.~M.; Ng,~K.~K. Physics of Semiconductor Devices; John
  Wiley \& Sons, 2006\relax
\mciteBstWouldAddEndPuncttrue
\mciteSetBstMidEndSepPunct{\mcitedefaultmidpunct}
{\mcitedefaultendpunct}{\mcitedefaultseppunct}\relax
\EndOfBibitem
\bibitem[Not()]{Note-2}
It is this error that is responsible for the experimental uncertainty in the
  position of the bands, and hence for the band offsets and the band gap
  extracted from the measurements. What is most strongly affected from this
  uncertainty are the intra-and interlayer exciton binding energies, which are
  are small quantities, obtained from the difference of much larger values
  (that is why the relative error on the extracted exciton binding energy is
  large).\relax
\mciteBstWouldAddEndPunctfalse
\mciteSetBstMidEndSepPunct{\mcitedefaultmidpunct}
{}{\mcitedefaultseppunct}\relax
\EndOfBibitem
\bibitem[Ugeda \latin{et~al.}(2014)Ugeda, Bradley, Shi, da~Jornada, Zhang, Qiu,
  Ruan, Mo, Hussain, Shen, Wang, Louie, and Crommie]{ugeda_giant_2014}
Ugeda,~M.~M.; Bradley,~A.~J.; Shi,~S.-F.; da~Jornada,~F.~H.; Zhang,~Y.;
  Qiu,~D.~Y.; Ruan,~W.; Mo,~S.-K.; Hussain,~Z.; Shen,~Z.-X.; Wang,~F.; Louie,~S.G.; Crommie,~M.F. Giant Bandgap Renormalization and Excitonic Effects in a Monolayer Transition
  Metal Dichalcogenide Semiconductor. \emph{Nat. Mater.} \textbf{2014},
  \emph{13}, 1091--1095\relax
\mciteBstWouldAddEndPuncttrue
\mciteSetBstMidEndSepPunct{\mcitedefaultmidpunct}
{\mcitedefaultendpunct}{\mcitedefaultseppunct}\relax
\EndOfBibitem
\bibitem[Zhang \latin{et~al.}(2016)Zhang, Chen, Huang, Wu, Li, Yao, Tersoff,
  and Shih]{Zhang2016}
Zhang,~C.; Chen,~Y.; Huang,~J.~K.; Wu,~X.; Li,~L.~J.; Yao,~W.; Tersoff,~J.;
  Shih,~C.~K. Visualizing Band Offsets and Edge States in Bilayer-Monolayer
  Transition Metal Dichalcogenides Lateral Heterojunction. \emph{Nat. Commun.}
  \textbf{2016}, \emph{6}, 1--6\relax
\mciteBstWouldAddEndPuncttrue
\mciteSetBstMidEndSepPunct{\mcitedefaultmidpunct}
{\mcitedefaultendpunct}{\mcitedefaultseppunct}\relax
\EndOfBibitem
\bibitem[Huang \latin{et~al.}(2016)Huang, Ding, Zhang, Chang, Shi, Li, Song,
  Zheng, Chi, Quek, and Wee]{Huang2016}
Huang,~Y.~L.; Ding,~Z.; Zhang,~W.; Chang,~Y.~H.; Shi,~Y.; Li,~L.~J.; Song,~Z.;
  Zheng,~Y.~J.; Chi,~D.; Quek,~S.~Y.; Wee,~A.T.S.; Gap States at Low-Angle
  Grain Boundaries in Monolayer Tungsten Diselenide. \emph{Nano Lett.}
  \textbf{2016}, \emph{16}, 3682--3688\relax
\mciteBstWouldAddEndPuncttrue
\mciteSetBstMidEndSepPunct{\mcitedefaultmidpunct}
{\mcitedefaultendpunct}{\mcitedefaultseppunct}\relax
\EndOfBibitem
\bibitem[Zhang \latin{et~al.}(2017)Zhang, Chuu, Ren, Li, Li, Jin, Chou, and
  Shih]{zhang_interlayer_2017}
Zhang,~C.; Chuu,~C.~P.; Ren,~X.; Li,~M.~Y.; Li,~L.~J.; Jin,~C.; Chou,~M.~Y.;
  Shih,~C.~K. Interlayer Couplings, Moiré Patterns, and 2D Electronic Superlattices in MoS$_{2}$/WSe$_{2}$ hetero-bilayers. \emph{Sci. Adv.}
  \textbf{2017}, \emph{3}, 1--8\relax
\mciteBstWouldAddEndPuncttrue
\mciteSetBstMidEndSepPunct{\mcitedefaultmidpunct}
{\mcitedefaultendpunct}{\mcitedefaultseppunct}\relax
\EndOfBibitem
\bibitem[Kang and Frisbie(2013)Kang, and Frisbie]{kang_pedagogical_2013}
Kang,~M.~S.; Frisbie,~C.~D. A Pedagogical Perspective on Ambipolar FETs. \emph{ChemPhysChem} \textbf{2013}, \emph{14},
  1547--1552\relax
\mciteBstWouldAddEndPuncttrue
\mciteSetBstMidEndSepPunct{\mcitedefaultmidpunct}
{\mcitedefaultendpunct}{\mcitedefaultseppunct}\relax
\EndOfBibitem
\bibitem[Mak \latin{et~al.}(2013)Mak, He, Lee, Lee, Hone, Heinz, and
  Shan]{Mak2013}
Mak,~K.~F.; He,~K.; Lee,~C.; Lee,~G.~H.; Hone,~J.; Heinz,~T.~F.; Shan,~J.
  Tightly Bound Trions in Monolayer MoS$_{2}$. \emph{Nat. Mater.} \textbf{2013},
  \emph{12}, 207--211\relax
\mciteBstWouldAddEndPuncttrue
\mciteSetBstMidEndSepPunct{\mcitedefaultmidpunct}
{\mcitedefaultendpunct}{\mcitedefaultseppunct}\relax
\EndOfBibitem
\bibitem[Bellus \latin{et~al.}(2015)Bellus, Ceballos, Chiu, and
  Zhao]{Bellus2015}
Bellus,~M.~Z.; Ceballos,~F.; Chiu,~H.~Y.; Zhao,~H. Tightly Bound Trions in
  Transition Metal Dichalcogenide Heterostructures. \emph{ACS Nano}
  \textbf{2015}, \emph{9}, 6459--6464\relax
\mciteBstWouldAddEndPuncttrue
\mciteSetBstMidEndSepPunct{\mcitedefaultmidpunct}
{\mcitedefaultendpunct}{\mcitedefaultseppunct}\relax
\EndOfBibitem
\bibitem[Rivera \latin{et~al.}(2016)Rivera, Seyler, Yu, Schaibley, Yan,
  Mandrus, Yao, and Xu]{rivera_valley-polarized_2016}
Rivera,~P.; Seyler,~K.~L.; Yu,~H.; Schaibley,~J.~R.; Yan,~J.; Mandrus,~D.~G.;
  Yao,~W.; Xu,~X. Valley-Polarized Exciton Dynamics in a 2D Semiconductor
  Heterostructure. \emph{Science (80-. ).} \textbf{2016}, \emph{351},
  688--691\relax
\mciteBstWouldAddEndPuncttrue
\mciteSetBstMidEndSepPunct{\mcitedefaultmidpunct}
{\mcitedefaultendpunct}{\mcitedefaultseppunct}\relax
\EndOfBibitem
\bibitem[Ross \latin{et~al.}(2017)Ross, Rivera, Schaibley, Lee-Wong, Yu,
  Taniguchi, Watanabe, Yan, Mandrus, Cobden, Yao, and Xu]{ross_interlayer_2017}
Ross,~J.~S.; Rivera,~P.; Schaibley,~J.; Lee-Wong,~E.; Yu,~H.; Taniguchi,~T.;
  Watanabe,~K.; Yan,~J.; Mandrus,~D.; Cobden,~D.; Yao,~W.; Xu,~X. Interlayer
  Exciton Optoelectronics in a 2D Heterostructure p–n Junction. \emph{Nano Lett.} \textbf{2017}, \emph{17}, 638--643\relax
\mciteBstWouldAddEndPuncttrue
\mciteSetBstMidEndSepPunct{\mcitedefaultmidpunct}
{\mcitedefaultendpunct}{\mcitedefaultseppunct}\relax
\EndOfBibitem
\bibitem[Wilson \latin{et~al.}(2017)Wilson, Nguyen, Seyler, Rivera, Marsden,
  Laker, Constantinescu, Kandyba, Barinov, Hine, Xu, and Cobden]{Wilson2017}
Wilson,~N.~R.; Nguyen,~P.~V.; Seyler,~K.; Rivera,~P.; Marsden,~A.~J.;
  Laker,~Z.~P.; Constantinescu,~G.~C.; Kandyba,~V.; Barinov,~A.; Hine,~N.~D. Xu,~X.; Cobden,~D.H. Determination of Band Offsets, Hybridization, and Exciton
  Binding in 2D Semiconductor Heterostructures. \emph{Sci. Adv.} \textbf{2017},
  \emph{3}, 1--8\relax
\mciteBstWouldAddEndPuncttrue
\mciteSetBstMidEndSepPunct{\mcitedefaultmidpunct}
{\mcitedefaultendpunct}{\mcitedefaultseppunct}\relax
\EndOfBibitem
\bibitem[Zhu \latin{et~al.}(2017)Zhu, Wang, Gong, Kim, Hone, and Zhu]{Zhu2017}
Zhu,~H.; Wang,~J.; Gong,~Z.; Kim,~Y.~D.; Hone,~J.; Zhu,~X.~Y. Interfacial
  {Charge} {Transfer} {Circumventing} {Momentum} {Mismatch} at
  {Two}-{Dimensional} van der {Waals} {Heterojunctions}. \emph{Nano Lett.}
  \textbf{2017}, \emph{17}, 3591--3598\relax
\mciteBstWouldAddEndPuncttrue
\mciteSetBstMidEndSepPunct{\mcitedefaultmidpunct}
{\mcitedefaultendpunct}{\mcitedefaultseppunct}\relax
\EndOfBibitem
\bibitem[Lee \latin{et~al.}(2014)Lee, Lee, van~der Zande, Chen, Li, Han, Cui,
  Arefe, Nuckolls, Heinz, Guo, Hone, and Kim]{lee_atomically_2014}
Lee,~C.-H.; Lee,~G.-H.; van~der Zande,~A.~M.; Chen,~W.; Li,~Y.; Han,~M.;
  Cui,~X.; Arefe,~G.; Nuckolls,~C.; Heinz,~T.~F.; Guo,~J.; Hone,~J.;, Kim,~P. Atomically
  Thin p–n Junctions with van der Waals Heterointerfaces. \emph{Nat.
  Nanotechnol.} \textbf{2014}, \emph{9}, 676--681\relax
\mciteBstWouldAddEndPuncttrue
\mciteSetBstMidEndSepPunct{\mcitedefaultmidpunct}
{\mcitedefaultendpunct}{\mcitedefaultseppunct}\relax
\EndOfBibitem
\bibitem[Miller \latin{et~al.}(1985)Miller, Chemla, Damen, Gossard, Wiegmann,
  Wood, and Burrus]{Miller1985}
Miller,~D. A.~B.; Chemla,~D.; Damen,~T.; Gossard,~A.; Wiegmann,~W.; Wood,~T.;
  Burrus,~C. Electric Field Dependence of Optical Absorption Near the Band Gap
  of Quantum-Well Structures. \emph{Phys. Rev. B} \textbf{1985}, \emph{32},
  1043--1060\relax
\mciteBstWouldAddEndPuncttrue
\mciteSetBstMidEndSepPunct{\mcitedefaultmidpunct}
{\mcitedefaultendpunct}{\mcitedefaultseppunct}\relax
\EndOfBibitem
\bibitem[Collins \latin{et~al.}(1986)Collins, Klitzing, and
  Ploog]{collins_photocurrent_1986}
Collins,~R.~T.; Klitzing,~K.~v.; Ploog,~K. Photocurrent Spectroscopy of Photocurrent Spectroscopy of GaAs/Al$_{x}$Ga$_{1-x}$As quantum wells in an electric field.
   \emph{Phys. Rev. B} \textbf{1986},
  \emph{33}, 4378--4381\relax
\mciteBstWouldAddEndPuncttrue
\mciteSetBstMidEndSepPunct{\mcitedefaultmidpunct}
{\mcitedefaultendpunct}{\mcitedefaultseppunct}\relax
\EndOfBibitem
\end{mcitethebibliography}
\end{document}